\documentclass[11pt]{article}

\usepackage[margin=1in]{geometry}
\usepackage[utf8]{inputenc}
\usepackage[T1]{fontenc}
\usepackage[numbers,sort&compress]{natbib}
\usepackage{hyperref}
\usepackage{url}
\usepackage{booktabs}
\usepackage{amsfonts}
\usepackage{amsmath}
\usepackage{amsthm}
\usepackage{graphicx}
\usepackage[table]{xcolor}
\usepackage{threeparttable}
\usepackage{pdfpages}

\definecolor{calmM}{HTML}{2E7D32}
\definecolor{calmMO}{HTML}{F9A825}
\definecolor{calmNM}{HTML}{C62828}

\newtheorem{theorem}{Theorem}
\newtheorem{proposition}{Proposition}

\title{When Coordination Is Avoidable:\\
A Monotonicity Analysis of Organizational Tasks}

\author{Harang Ju\\
Carey Business School, Johns Hopkins University\\
Baltimore, MD 21202\\
\texttt{harang@jhu.edu}}

\date{}

\begin{document}

\maketitle

\begin{abstract}
Organizations devote substantial resources to coordination, yet which tasks actually require it for correctness remains unclear. The problem is acute in multi-agent AI systems, where coordination cost is directly measurable and can exceed the cost of the work itself. Distributed systems theory provides a precise criterion: coordination is required when a task specification is non-monotonic, meaning that as histories grow, new information can invalidate prior conclusions. Here we show that Thompson's classic taxonomy of interdependence maps to that criterion, yielding a decision rule for when coordination is required for correctness. We formalize the correspondence in a bridge theorem, apply the rule to 65 workflows from the American Productivity \& Quality Center (APQC) and, with a calibrated large language model (LLM), 13,417 Occupational Information Network (O*NET) tasks, and illustrate it in multi-agent AI simulations. Under our decompositions, 74\% of workflows and 42\% of O*NET tasks are monotonic, implying that up to 24--57\% of coordination spending is unnecessary for correctness.
\end{abstract}


\section*{Introduction}

Coordination is a fundamental cost of organization. Meetings, approval chains, status updates, and cross-functional reviews all exist because organizations assume that coordination is necessary. Yet no formal criterion distinguishes the coordination that is necessary for correctness from the coordination that is not. The cost of this ambiguity is substantial and growing, especially with the recent proliferation of AI agents. In multi-agent AI systems, which inherit their coordination topology from the organizations they assist, cost grows super-linearly with group size \cite{chen2025scaling}. Because these systems make coordination costs directly countable in tokens, they also provide a clean empirical setting for testing theories of coordination necessity. The standard response is to optimize coordination by improving communication, streamlining processes, or adding coordination technology. Optimization, however, assumes that coordination is necessary. What if much of it is not?

Whether coordination is necessary depends on the structure of the task. Thompson's \cite{thompson1967organizations} interdependence taxonomy classifies tasks by the coordination demands their structure imposes. Pooled tasks contribute independently to a common outcome and need only uniform rules. Sequential tasks pass output from one unit to the next and require planning and scheduling. Reciprocal tasks require mutual adjustment among units, the most costly coordination form. Thompson observed that pooled and sequential interdependence dominate, with reciprocal interdependence concentrating at critical integration points. This taxonomy remains foundational in organizational theory. Subsequent work has drawn sharper distinctions among task interdependence, agent interdependence, epistemic interdependence, and the fit between interdependence structures and coordination mechanisms \cite{puranam2012epistemic, puranam2018microstructure}. Here, we focus on one formal dimension: when coordination is required for correctness under a specified task representation.

Distributed computing contributes a formal criterion on this dimension. The CALM theorem (Consistency As Logical Monotonicity) proves that a computation can execute without coordination if and only if its specification is monotonic, meaning that as admissible histories expand, previously valid conclusions need not be retracted \cite{hellerstein2020calm, ameloot2013relational}. Non-monotonic computations, where new information can invalidate prior conclusions, provably require coordination for correctness. The theorem has been applied extensively to databases \cite{bailis2015coordination} and recently unified with other coordination criteria over Lamport histories \cite{hellerstein2026coordination}, but not to organizational task analysis. CALM governs correctness, not broader performance objectives. A monotonic task may still benefit from coordination for speed, stylistic consistency, or other goals, but correctness does not require it.

In this paper, we map Thompson's taxonomy to CALM's monotonicity criterion. The Thompson-CALM Bridge Theorem (Theorem~\ref{thm:bridge}) shows that pooled interdependence maps to monotone specifications and sequential interdependence maps to monotone specifications under causal ordering. The reciprocal case depends on feedback type. Reciprocal tasks become non-monotone when feedback retracts or constrains previously admissible outcomes, but can remain monotone when agents only add to one another's output. The Feedback Boundary (Proposition~\ref{prop:feedback}) extends the same distinction to sequential tasks with feedback, clarifying when feedback preserves monotonicity and when it introduces non-monotonic revision. The Coordination Tax (Equation~\ref{eq:ceiling}) then quantifies the upper bound on avoidable coordination cost under a regime that coordinates all tasks uniformly.

We apply this framework to two complementary corpora. We classify 65 enterprise workflows across all 13 categories of the APQC Process Classification Framework \cite{apqc2024pcf} and find that 74\% are monotonic under our decompositions. We extend this classification to 13,417 occupational tasks from the O*NET 29.1 database \cite{onet2024} spanning 22 Standard Occupational Classification (SOC) major groups using a calibrated LLM annotator \cite{eloundou2024gpts, massenkoff2026labor}, and classify 42\% as monotonic under multi-agent decomposition. These rates imply that up to 24--57\% of coordination spending is unnecessary for correctness under uniform coordination. Multi-agent AI simulations then illustrate the distinction experimentally across three models, where coordinated and uncoordinated conditions differ in whether agents can revise or reconcile one another's outputs and validity is assessed against pre-specified criteria.

Taken together, these results connect several lines of organizational theory. The Bridge Theorem contributes a formal criterion for a question that Thompson's \cite{thompson1967organizations} verbal classification left implicit and that Malone and Crowston's \cite{malone1994interdisciplinary} program of characterizing dependencies left open, namely when coordination is necessary for correctness at all. The contribution is complementary to contemporary organization design work, which emphasizes that actual coordination requirements also depend on who knows what, who influences whom, and which coordination mechanism is deployed \cite{puranam2012epistemic, puranam2018microstructure}. Task topology can inform coordination intensity for correctness-critical work, whether in human organizations or AI systems.

\section*{Theory}

\subsection*{Theoretical Bridge Between Interdependence and Monotonicity}

Coordination is required for correctness when local actions can invalidate previously admissible outcomes. Distributed systems theory formalizes this condition using specification monotonicity: a task is monotonic if extending the execution history cannot invalidate previously admissible conclusions. Organizational theory classifies tasks by patterns of interdependence instead. We show that these two traditions align except for the reciprocal case, where alignment is conditional on feedback type.

Establishing the correspondence requires three definitions. We define correctness as the property that a system's outputs satisfy a specified constraint under all admissible execution orders. We define coordination as any mechanism that enforces ordering or synchronization to prevent inconsistent intermediate states. A task requires coordination if correctness cannot be guaranteed under asynchronous, unordered execution. Formal definitions appear in SI Text S1.

Thompson distinguishes pooled, sequential, and reciprocal interdependence. Pooled tasks aggregate independent contributions, and their outputs combine via a join-semilattice \cite{shapiro2011crdt}. Sequential tasks require ordered handoffs along Lamport's happened-before order \cite{lamport1978time}, where each agent's computation is non-retractive. Reciprocal tasks involve mutual adjustment among agents, but reciprocal structure alone does not settle monotonicity. The relevant distinction is whether feedback merely adds information or whether it retracts or constrains previously admissible outcomes.

CALM provides the complementary criterion from distributed systems: a specification admits coordination-free execution if and only if it is monotonic \cite{ameloot2013relational, hellerstein2020calm, hellerstein2026coordination}. Monotonicity is orthogonal to recursion or feedback as such. Recursive or feedback-rich tasks may still be monotonic if each history extension only enlarges what can be concluded. By contrast, specifications that require retraction, negation, or other shrinking of admissible outcomes require coordination. The following theorem states the bridge we establish.

\begin{theorem}[Thompson-CALM Bridge]\label{thm:bridge}
Under deterministic task evaluation, reliable message delivery, and eventual consistency in state propagation, let $T$ be a multi-agent task. Then:
\begin{enumerate}
    \item[(a)] Pooled interdependence $\implies$ the task specification is monotone (coordination-free).
    \item[(b)] Sequential interdependence $\implies$ the task specification is monotone under causal ordering (coordination-free with causal delivery).
\end{enumerate}
\end{theorem}

Both cases hinge on non-retractive composition. For (a), sub-task monotonicity and the join-semilattice property guarantee that aggregate output only grows. For (b), the composition of non-retractive functions is non-retractive, so causal delivery suffices. Reciprocal interdependence alone does not determine monotonicity, but feedback that retracts or constrains previously admissible outcomes makes the specification non-monotone by definition and therefore requires coordination. Full proofs appear in SI Text S2.

The Bridge Theorem classifies pure sequential tasks as monotonic under non-retractive causal composition, but many real workflows contain feedback loops such as code review, quality inspection, and editorial passes. The Feedback Boundary resolves whether feedback pushes a task across the monotonicity boundary.

\begin{proposition}[Feedback Boundary]\label{prop:feedback}
Let $T$ be a sequential task with feedback from a downstream agent to an upstream agent.
\begin{enumerate}
    \item[(a)] If the feedback is additive\footnote{Additive and retractive are informal labels for monotonic and non-monotone feedback, respectively, where new information only refines or else invalidates prior output; e.g., tightening a bound from ``at least 5'' to ``at least 6'' preserves the earlier conclusion, so it is monotone even if plain English would call it a retraction.}, so the upstream agent incorporates new information without retracting prior output, the specification remains monotone.
    \item[(b)] If the feedback is retractive, so the upstream agent revises or replaces prior output, the specification is non-monotone.
\end{enumerate}
\end{proposition}

Additive feedback preserves the non-retractive property established in Theorem~\ref{thm:bridge}(b), whereas retractive feedback violates it. This distinction maps onto Argyris and Sch\"on's \cite{argyris1978organizational} boundary between single-loop refinement and double-loop revision.

\subsection*{The Coordination Tax}

The Bridge Theorem identifies which tasks need coordination under the paper's formal assumptions. The Coordination Tax quantifies how much coordination cost is avoidable when that classification is ignored. Let $\mathcal{P}$ be a portfolio of $n$ tasks with non-monotonic fraction $f \in [0,1]$ and coordination cost multiplier $c > 1$. If every task receives uniform coordination, the unnecessary fraction of total spending is
\begin{equation}\label{eq:ceiling}
    T(f, c) = \frac{(1-f)(c-1)}{c}
\end{equation}
Equation~(\ref{eq:ceiling}) is an upper bound on avoidable coordination cost under a regime in which all tasks receive uniform coordination regardless of structural necessity. It is not a direct measure of realized waste in organizations, which may already use differentiated coordination mechanisms such as standardization, modularization, routines, authority, and mutual adjustment.

The derivation is straightforward. Uniform cost $nc$ minus selective cost $n(1{-}f{+}fc)$ equals $n(1{-}f)(c{-}1)$; dividing by $nc$ gives $T$. As $f \to 0$ (no tasks need coordination), $T \to (c{-}1)/c$; as $f \to 1$ (all tasks need coordination), $T \to 0$.

\section*{Results}

We classify two task corpora under the Bridge Theorem and illustrate the decision rule in multi-agent AI simulations.

\subsection*{Most Enterprise Workflows Are Monotonic (74\%)}

Our first corpus draws from the APQC Process Classification Framework \cite{apqc2024pcf}, the most widely used enterprise process taxonomy (maintained since 1992). We classify 65 workflows at Levels~2--3 (team-level workflows with 2--5 agents), enumerating all eligible processes as an exhaustive census across all 13 APQC categories (see Materials and Methods).

Applying the Bridge Theorem to this corpus, we find that 48 of 65 tasks (73.8\%) are monotonic (Table~\ref{tab:prevalence}), of which 39 are fully coordination-free and 9 require only causal ordering. The remaining 17 (26.2\%) require full coordination. These labels reflect the task decompositions and heuristic decision rules stated in Materials and Methods, with borderline cases defaulting to non-monotonic.

\begin{table}[ht!]
\centering
\caption{\textbf{Most enterprise tasks (APQC) do not require coordination for correctness.} CALM monotonicity classification of 65 enterprise tasks across all 13 APQC categories. M: coordination-free; M-O: coordination-free under causal ordering; NM: coordination required.}
\label{tab:prevalence}
{\small
\begin{tabular}{lcccc}
\toprule
\textbf{APQC Category} & \textbf{M} & \textbf{M-O} & \textbf{NM} & \textbf{\%\ Monotonic} \\
\midrule
Risk \& Compliance & 3 & 1 & 0 & 100\% \\
Vision \& Strategy & 3 & 1 & 1 & 80\% \\
Products \& Services & 3 & 1 & 1 & 80\% \\
Marketing \& Sales & 3 & 1 & 1 & 80\% \\
Service Delivery & 3 & 1 & 1 & 80\% \\
Customer Service & 3 & 1 & 1 & 80\% \\
Assets & 3 & 0 & 1 & 75\% \\
External Relations & 3 & 0 & 1 & 75\% \\
Business Capabilities & 3 & 0 & 1 & 75\% \\
Human Capital & 4 & 0 & 2 & 67\% \\
Information Technology & 3 & 1 & 2 & 67\% \\
Physical Products (info) & 3 & 0 & 2 & 60\% \\
Financial Resources & 2 & 2 & 3 & 57\% \\
\midrule
\textbf{Total} & \textbf{39} & \textbf{9} & \textbf{17} & \textbf{73.8\%} \\
\bottomrule
\end{tabular}
}
\end{table}

Breaking the result down by APQC category, we find systematic variation (Table~\ref{tab:prevalence}). Risk \& Compliance ranks highest at 100\% because its tasks consist primarily of independent analyses and reports that merge without conflict, five categories share 80\%, and Financial Resources ranks lowest at 57\% because most of its tasks involve allocating shared budgets or credit across competing claims.

To understand what drives the 26\% that require coordination under this framework, we examine the non-monotonic tasks individually and find that all 17 involve allocation of shared finite resources (budget, headcount, capacity, inventory). The pattern echoes Thompson's observation that reciprocal interdependence clusters at critical integration points \cite{thompson1967organizations}. Shared finite resources introduce negation or retractive constraint, which breaks monotonicity. Under these decompositions, the bulk of enterprise workflows can execute correctly without coordination.

Because these labels may depend on decomposition choices, we also performed a sensitivity analysis on a small sample of ten APQC tasks (five M/M-O and five NM, spread across APQC categories) reclassified under alternative reasonable assumptions about timing, shared-resource scope, or review semantics (SI Text S9). Seven of ten retained their monotonicity status, and three crossed the monotonicity boundary under the alternate decomposition. These boundary crossings show that decomposition choices can sometimes introduce or remove non-monotonicity by changing where coordination constraints are represented. Of the seven stable tasks, two also shifted from ordering-only to fully monotone, which does not affect the monotonicity rate. Taken together, most tasks retain the same monotonicity status under reasonable decomposition variation, and the 74\% monotonicity rate remains decomposition-dependent rather than an invariant property of the tasks.

\subsection*{Nearly Half of Occupational Tasks Are Monotonic (42\%)}

\begin{figure}[t]
    \centering
    \includegraphics[width=\textwidth]{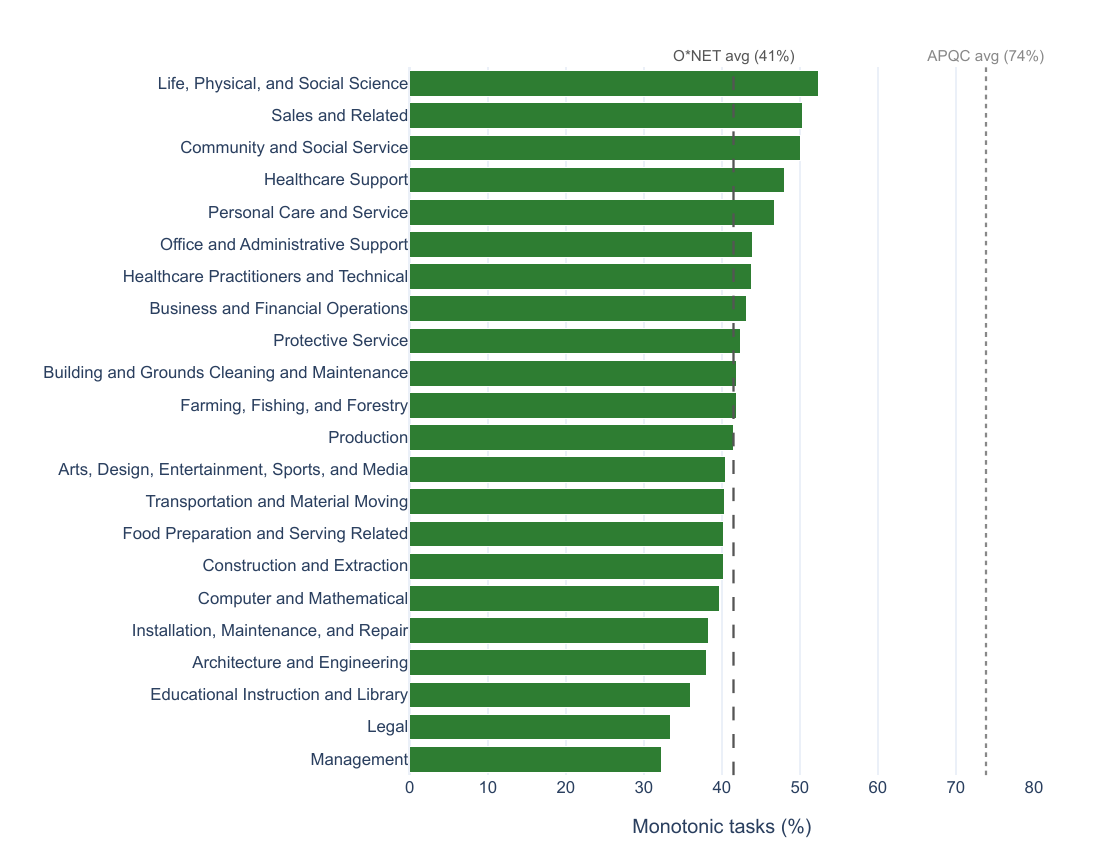}
    \caption{\textbf{Monotonic tasks appear in every occupational category (O*NET).} Bars show monotonicity prevalence across 22 SOC major groups (13,417 O*NET tasks); every group exceeds 30\%. Dashed line marks the O*NET average (41\%); dotted line marks the APQC enterprise average (74\%).}
    \label{fig:onet}
\end{figure}

To test whether monotonicity prevalence extends beyond enterprise workflows, we apply a calibrated LLM classifier ($\kappa \geq 0.88$ against author labels on APQC; see Materials and Methods) to 13,417 core task statements from the O*NET 29.1 occupational database \cite{onet2024} spanning 22 SOC major groups (Figure~\ref{fig:onet}). Of these, 5,564 (41.5\%, 95\% CI [40.6, 42.3]) are monotonic. The rate is lower than APQC's 74\%, which reflects the difference between team-level workflows (APQC) and individual work activities (O*NET). APQC processes decompose naturally into multi-agent sub-tasks, whereas O*NET tasks describe what a single worker does. The O*NET classification therefore characterizes the inferred coordination structure of each task under multi-agent decomposition.

Figure~\ref{fig:onet} shows the breakdown by SOC group. Monotonic tasks appear in every group, from Life, Physical, and Social Science (52\%) to Management (32\%). Groups dominated by independent observation and analysis (sciences, sales, healthcare support) rank highest, while Management ranks lowest because it concentrates resource allocation, staffing, and cross-unit negotiation. Full per-group results with 95\% Wilson confidence intervals appear in SI Table S4. Under this decomposition, the scope for coordination avoidance extends beyond enterprise workflows to the broader occupational landscape.

\subsection*{Estimating the Coordination Tax}

With both corpora classified, we can compute the Coordination Tax using Equation~\ref{eq:ceiling}, $T(f,c)=(1-f)(c-1)/c$, which assumes that all tasks receive uniform coordination regardless of structural necessity. The non-monotonic fraction is $f = 17/65 \approx 0.26$ for APQC (Table~\ref{tab:prevalence}) and $f = 7{,}852/13{,}417 \approx 0.58$ for O*NET (SI Table S4). Substituting these fractions and the simulation-measured overhead ratios ($c = 2.3$--$4.4\times$, SI Table S7) gives an APQC-derived tax of $T \approx 42$--$57\%$ and an O*NET-derived tax of $T \approx 24$--$32\%$.

The gap between corpora reflects the unit of analysis: APQC describes team-level workflows where coordination cost concentrates, while O*NET describes individual work activities. Even at the most conservative estimate (O*NET, $c = 2.3\times$), the formula places nearly a quarter of coordination cost in the avoidable range. These quantities are upper bounds under stated assumptions, not direct estimates of realized organizational inefficiency, and organizations that already differentiate coordination across tasks may realize only a fraction.

\subsection*{Simulations Illustrate the Monotonicity Boundary}

To illustrate the framework's prediction in a multi-agent AI setting, we run simulations where coordination can be cleanly toggled and costs are directly measurable in tokens (see Materials and Methods). In the coordinated condition, an orchestrator plans assignments, passes intermediate outputs between agents, and reconciles conflicts where needed. In the uncoordinated condition, agents receive only their sub-task descriptions and their outputs are concatenated without review. We select ten tasks spanning the monotonicity spectrum and run each under both conditions across three models (GPT-4.1 mini, Claude Haiku 4.5, Claude Sonnet 4.5; 10 repetitions at temperature~0). Validity is assessed against pre-specified task criteria by an LLM judge, with programmatic checks for the resource-constrained tasks.

\begin{table}[ht]
\centering
\caption{\textbf{Uncoordinated validity rate in simulations across three models.} Coordinated condition substantially recovers non-monotonic validity across models; full per-model rates appear in SI Text S8. Bold = 0\%. M = monotonic, M-O = monotonic with ordering, NM = non-monotonic.}
\label{tab:demonstration_multimodel}
\small
\begin{tabular}{llccc}
\toprule
Task & Type & GPT-4.1 mini & Haiku 4.5 & Sonnet 4.5 \\
\midrule
Strategy pillars & M & 100\% & 100\% & 100\% \\
Feature specs & M & 100\% & 100\% & 100\% \\
Marketing content & M & 100\% & 100\% & 90\% \\
Security audit & M & 100\% & 100\% & 100\% \\
\midrule
Stage-gate & M-O & 100\% & 100\% & 100\% \\
Ticket escalation & M-O & 100\% & 60\% & 70\% \\
\midrule
Budget alloc. & NM & \textbf{0\%} & \textbf{0\%} & \textbf{0\%} \\
Backlog sprint & NM & \textbf{0\%} & \textbf{0\%} & \textbf{0\%} \\
Production sched. & NM & \textbf{0\%} & \textbf{0\%} & \textbf{0\%} \\
Headcount alloc. & NM & \textbf{0\%} & \textbf{0\%} & \textbf{0\%} \\
\bottomrule
\end{tabular}
\end{table}

The prediction holds across all three models (Table~\ref{tab:demonstration_multimodel}). For GPT-4.1 mini, all six monotonic-or-ordered tasks achieve 100\% validity in both conditions, while every non-monotonic task drops to 0\% validity without coordination. The Claude models replicate the boundary result: uncoordinated non-monotonic tasks fail at 0\% across all four tasks, while monotonic tasks remain near ceiling. Remaining variance on ordering-only tasks tracks model capability rather than violating the framework. Validity is binary in these runs, whereas organizational failures more often manifest as gradual degradation or rework. Coordinated runs consume 2.3--4.4 times more tokens (SI Table S7) and substantially recover non-monotonic validity.

\section*{Discussion}

We show that for a substantial fraction of tasks, coordination is not required for correctness. The Bridge Theorem identifies which tasks those are through a partial correspondence between Thompson's taxonomy and CALM's monotonicity criterion, and the Coordination Tax bounds the avoidable coordination cost under a regime that coordinates all tasks uniformly. Across two corpora, 42--74\% of tasks are monotonic under our decompositions, and 24--57\% of coordination cost would be avoidable under uniform-coordination assumptions. A sensitivity analysis on APQC tasks shows that plausible alternative decompositions can move some tasks across the monotonicity boundary, especially when review is revisionary or shared resources are pre-partitioned, but most tasks retained their monotonicity status. Simulations illustrate the distinction operationally in a controlled multi-agent AI setting.

The Bridge Theorem clarifies one dimension of interdependence that Thompson \cite{thompson1967organizations} highlighted but did not formalize, namely whether extending the task history can invalidate previously admissible outcomes. Under this criterion, pooled and sequential structures align naturally with monotone specifications, whereas retractive or constraining feedback makes the specification non-monotone and reciprocal structure alone does not determine monotonicity. This partial correspondence shows that reciprocal feedback need not be non-monotone.

The monotone form of reciprocal feedback is additive feedback, which can support coordination-free feedback loops even though retractive or constraining feedback remains non-monotone. Recognizing additive feedback as monotone expands the space of coordination-free specifications and points to a broader question for organization science. One possible reason this case has received less attention is that many organizational feedback processes are studied in settings where feedback revises, approves, constrains, or reallocates prior work. Future work can ask when monotone feedback is practically achievable, and whether recognizing it can inform organizational and AI workflows that accumulate information without invalidating prior outputs.

Specification monotonicity also responds to long-standing critiques of Thompson. Victor and Blackburn \cite{victor1987interdependence} argued that Thompson's typology mixes structural, behavioral, and evaluative dimensions of interdependence, and McCann and Ferry \cite{mccann1979interdependence} called for operational measures beyond categorical labels. Specification monotonicity is one such semantic property, and it cuts across Thompson's categories. A reciprocal task with additive feedback can be monotonic, whereas a sequential task with retractive revision can be non-monotonic. Because the criterion is operationally testable by asking whether history extension can invalidate previously admissible conclusions, it speaks directly to both critiques.

This contribution is complementary to contemporary organization science. Task interdependence is distinct from agent interdependence, which concerns how agents' actions are coupled through learning, influence, or relational contracts, and from epistemic interdependence, which captures the distribution of knowledge across agents independent of task structure \cite{puranam2012epistemic}. A microstructural perspective further emphasizes that coordination needs depend on authority, influence, and the fit between interdependence structures and coordination mechanisms \cite{puranam2018microstructure}. Monotonicity is a property of the task specification, including epistemic limits, agent coupling, incentives, and mechanism constraints. For organization scholars, the contribution is a testable semantic criterion that cuts across existing typologies rather than a new theory of organizations.

The Bridge Theorem also advances the research program that Malone and Crowston \cite{malone1994interdisciplinary} outlined. They defined coordination as ``managing dependencies among activities'' and proposed that progress would come from characterizing dependency types and the coordination processes needed to manage them. The Bridge Theorem addresses a prior question: given a dependency type and a task specification, is coordination necessary at all for correctness? In that sense, it supplies a formal criterion that their framework did not provide.

The Coordination Tax isolates a category of coordination cost that is avoidable for correctness under a uniform-coordination regime. It does not imply that observed coordination in organizations is wasteful by that amount, because organizations often deploy differentiated mechanisms such as standardization, routines, authority, or modularization. The formula instead identifies the gap between a world that coordinates everything uniformly and a world that coordinates only when required for correctness. Baldwin and Clark's \cite{baldwin2000design} account of modularity fits naturally here. Modularity can reduce the non-monotonic share of a task portfolio by localizing retractive dependencies within a thinner integration layer.

These findings carry practical implications for multi-agent systems, human or artificial. Pooled tasks can often use fire-and-forget execution with merge by join. Sequential tasks can often rely on causal delivery without full mutual adjustment. Only the non-monotonic subset requires conflict resolution, synchronization, or other stronger coordination mechanisms. For AI systems in particular, this suggests that agent topology should mirror task dependency structure rather than default managerial hierarchy. For multi-agent system designers, this is also an invitation to draw on organization-design scholarship rather than around it, treating monotonicity as one input into a broader design problem.

The framework and evidence have several limitations. First, CALM guarantees correctness, not broader performance objectives, so independently written sections may be valid yet stylistically inconsistent, slow, redundant, or poorly prioritized. Second, misclassification risk is asymmetric: a false monotonic label endangers correctness, whereas a false non-monotonic label mainly adds unnecessary coordination cost. Third, the monotonicity labels are decomposition-dependent rather than invariant properties of the tasks, and alternative assumptions about timing, shared-resource scope, or review semantics can move some tasks across the boundary. Fourth, the simulations use binary validity outcomes in a controlled AI-agent sandbox rather than field observations of organizations, whereas real organizational failures often appear as delays, rework, partial inconsistency, or degraded quality rather than clean invalidity. Finally, LLM-assisted classification, as deployed here, is a calibrated screening tool rather than a substitute for verification, and the resulting labels may vary across prompts and model implementations.

A few directions would extend this work. First, a neurosymbolic approach would give stronger certainty about any particular task by coupling an LLM proposer with a formal verifier and enforcing the monotone protocol through implementation. Second, field observations of organizations would test how the monotonicity boundary manifests in realized workflow behavior, complementing the controlled AI-agent simulations reported here. Third, epistemic interdependence could be connected more directly to distributed-systems accounts of common knowledge, including the impossibility result of Halpern and Moses and recent complexity extensions to CALM \cite{halpern1990knowledge, hellerstein2026determinations}. Together, these would sharpen the framework from a theoretical upper bound under stated assumptions into directly testable claims about per-task correctness and observed coordination.

The broader implication is that whether a task requires coordination for correctness can be treated as a computable property of a formal task representation, providing a principled starting point for allocating coordination intensity. The strongest claim supported by the present evidence is that the framework identifies a theoretical upper bound on avoidable coordination under specified assumptions, not that the reported percentages represent realized inefficiency in any particular organization. As organizations and AI systems grow more complex, distinguishing correctness-critical coordination from coordination adopted by convention may become as important as improving coordination itself.

\section*{Materials and Methods}

\paragraph{APQC corpus construction.}
We draw our primary task corpus from version 7.4 of the APQC Process Classification Framework (Cross-Industry) \cite{apqc2024pcf}, the most widely used enterprise process taxonomy. We define tasks at Levels~2--3 (2--5 agents, team-level workflows). A process qualifies if it decomposes into 2+ agent sub-tasks, involves information processing, produces a defined output, and spans industries. We enumerate all eligible processes as an exhaustive census, yielding 65 tasks across all 13 APQC categories (SI Table S3). For each task, we classify interdependence structure using Thompson's taxonomy, apply the Bridge Theorem to predict CALM classification, and verify against five heuristic tests (SI Text S3). Borderline cases default to non-monotonic. These labels are best understood as structured proxy judgments about monotonicity, not direct formal verification of every workflow. We are careful to distinguish the theoretical construct (specification monotonicity under history growth) from the empirical proxy (structured judgments under a stated decomposition).

\paragraph{O*NET classification procedure.}
To scale beyond the 65 APQC tasks, we calibrate an LLM classifier against the author-classified APQC corpus. We use GPT-4.1 mini and Claude Sonnet 4.5, each under two prompting conditions (3 runs at temperature~0), achieving $\kappa \geq 0.88$ on the binary monotonic-versus-non-monotonic distinction (SI Table S5). Every misclassification labels a monotonic task as non-monotonic. We apply the calibrated classifier (GPT-4.1 mini, heuristic condition) to 13,417 core task statements from the O*NET 29.1 occupational database \cite{onet2024} spanning 22 SOC major groups, following Eloundou et al.\ \cite{eloundou2024gpts}, who used GPT-4 to annotate O*NET tasks for LLM exposure. Their theoretical exposure measures have since been empirically validated against observed LLM usage (97\% of observed Claude-usage tasks fall in categories Eloundou rated as theoretically feasible \cite{massenkoff2026labor}). The resulting labels are inferred under the paper's decomposition assumptions and should not be read as direct measures of occupational coordination in practice.

\paragraph{Simulation design.}
Ten tasks span the monotonicity spectrum (four monotonic, two requiring ordering only, four non-monotonic) under two conditions. In the coordinated condition, an orchestrator LLM plans assignments, passes intermediate outputs, reviews for consistency, and reconciles conflicts for the non-monotonic tasks. In the uncoordinated condition, agents receive only their sub-task descriptions and outputs are concatenated mechanically, with no review or reconciliation layer. An LLM-as-judge evaluator reasons step-by-step against pre-specified validity criteria before issuing a VALID or INVALID verdict. For the resource-constrained tasks, we supplement the judge with programmatic checks as reported in SI Text S8. We interpret these simulations as controlled illustrations in an AI setting rather than as direct evidence about organizational field behavior. We run 10 repetitions at temperature~0 across three models (GPT-4.1 mini, Claude Haiku 4.5, Claude Sonnet 4.5). Full task specifications, validity criteria, and token cost ratios appear in SI Text S8.

\paragraph{Data availability.}
The APQC Process Classification Framework is available at \url{https://www.apqc.org/process-frameworks} \cite{apqc2024pcf}. The O*NET 29.1 database is publicly available at \url{https://www.onetcenter.org/database.html} \cite{onet2024}. All classification data and simulation code are available at \url{https://github.com/harangju/coordination-avoidability} \cite{ju2026coordination}.

\paragraph{Author contributions.}
H.J.\ designed research; performed research; analyzed data; and wrote the paper.

\paragraph{Competing interests.}
The author declares no competing interests.

\bibliographystyle{unsrt}
\bibliography{references}

\clearpage
\includepdf[pages=-]{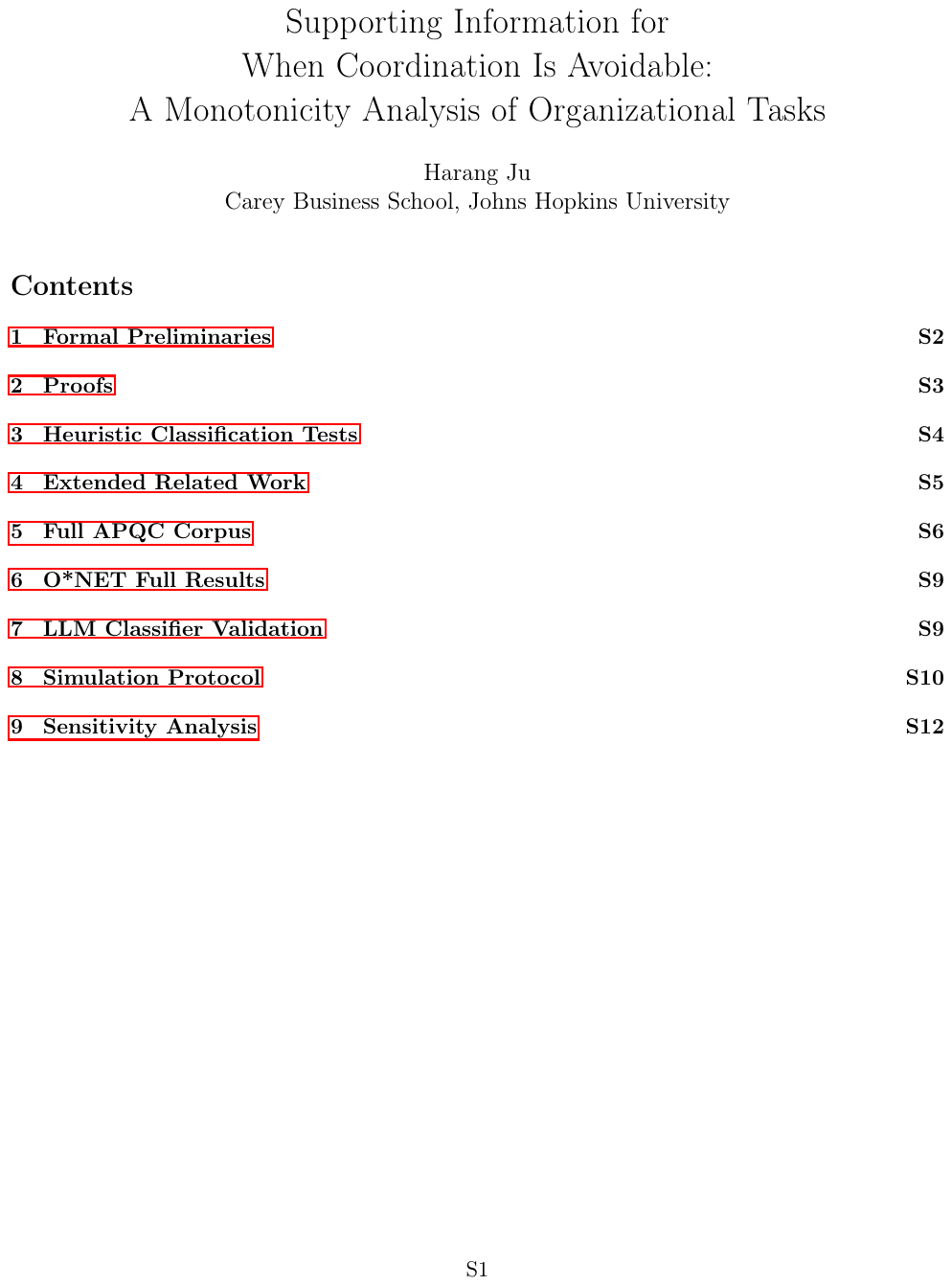}

\end{document}